\documentclass[12pt]{iopart}
\usepackage{graphicx}
\usepackage{iopams}
\eqnobysec
\begin{document}

\title{Analogy of Grounded Conducting Sphere Image Problem with Mirror-Optics}

\author{Kolahal Bhattacharya}

\address{Department of High Energy Physics, Tata Institute 
of Fundamental Research, Homi Bhabha Road, Mumbai, India}
\ead{kolahalb@tifr.res.in}

\begin{abstract}
We show that in the grounded conducting sphere image problem, all the necessary informations 
about the image charge can be found from a mirror equation and a magnification formula. Then, 
we propose an method to solve the image problem for an extended charge distribution near a 
grounded conducting sphere.
\end{abstract}

\submitto{EJP}
\maketitle

\section{Introduction}
Image problems in electrostatics refer to boundary-value problems that specify some charge 
distribution in a volume bounded by a closed surface and also specify the potential on the 
same surface. However, the problem is constructed such that the potential specified on the 
surface cannot be due to the known charge distribution only. The problem is, then, to find 
the general potential function in the volume where the known charge resides. To solve this 
problem, some other charge distribution outside the boundary is imagined such that at each 
point on the surface, the potentials due to inside and outside charges sum up to the value 
specified in the problem. The outside charge is called the image charge. It turns out that 
in some typical boundary value problems the image method works elegantly: a unique general 
potential function can be constructed.

For example, let us consider a point charge $q$ at a distance $d$ above an infinite grounded 
conducting plane. Because the potential must be zero everywhere on this plane, the required 
potential $\Phi$ above the plane (where $q$ resides), is the same as the sum of the potentials 
produced by $q$ (above the plane) and an image charge $-q$ placed at a distance $2d$ directly 
below $q$, inside the plane. Effectively, the image charge is like the optical image of the 
real charge except for the fact that a plane mirror does not need to be infinite to form optical 
image. We will try to explore this similarity in this article.

We will start by looking at the grounded conducting sphere problem (Figure 1). A charge $q$ 
is placed in front of a grounded conducting sphere ($\Phi=0$ on the spherical surface). The 
potential function at any point $\bf{r}$ in the region where $q$ resides is asked. We apply 
the image method to solve for the potential function. The image charge effectively replaces 
the charge induced on the sphere. Thus, $\Phi(\bf{r}) $ can be calculated as if there is no 
sphere and only the real charge $q$ and image charge $q'$ are there, provided that the position and 
magnitude of $q'$ are chosen such that $\Phi$ is zero on the surface originally occupied by 
the sphere. Then, the only job is to find value and position of the image charge. The results are 
already known and available in the standard texts in References. 

A little manipulation of the results will lead to not-so-well-known observations with great 
similarity with the standard results in geometrical optics. This will put more light on the 
existing theory of electrostatic images. It will also enable us to handle the image problem 
of an extended charged body, placed before a grounded conducting sphere. The problem is non-
trivial as the image charge is distorted (like optical image produced in a spherical mirror); 
and hence, the direct calculation of potential function is difficult. The usual approach is 
to integrate the Green's function of the generic point charge problem over the volume of the 
given charge distribution. We propose a more intuitive method that leads to the same formula.

\section{Grounded Conducting Sphere Image Problem}

\subsection{Standard results of grounded conducting sphere}

We consider the grounded conducting sphere image problem (Figure 1). According to reference 
[1], at some field point $\bf{r}$, the potential is the same as that produced by the charge 
$q$ at $S$ (such that $OS=y$) plus that by the image charge $q'=-\frac{aq}{y}$ placed at $I
$ (so that $OI=y'=\frac{a^2}{y}$). To appreciate the similarity of the present problem with 
light reflection off a spherical mirror, we first recall the standard results from optics.

\subsection{Spherical convex mirror in optics}
\begin{figure}[h]
\begin{center}
\scalebox{0.60}{\includegraphics{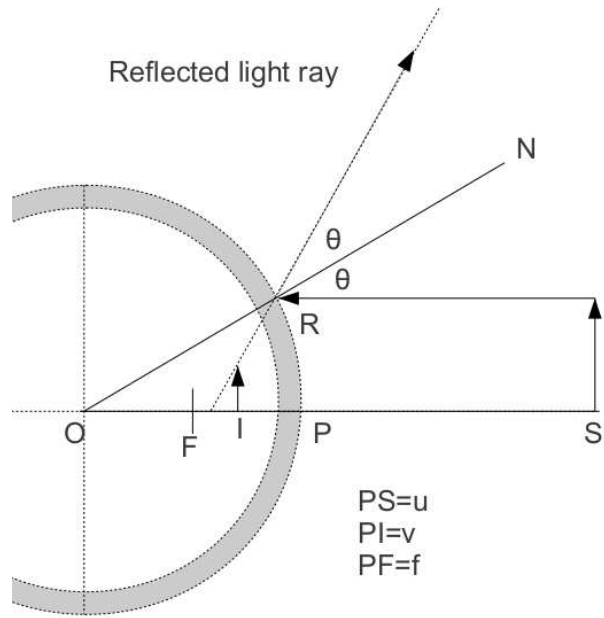}}
\caption{\label{fig:1} Reflection from a spherical mirror in optics}
\end{center}
\end{figure}
Light ray emitted from the object at $S$ is reflected off a convex spherical surface (Figure 
1) and the image forms at $I$. $F$ is the focus and $O$ is the centre of curvature. Then, we 
know that the object distance $PS=u$, the image distance $PI=v$, and the focal length $PF=f$ 
(distance from the pole where image forms if the object is at infinity: $u\rightarrow\infty$) 
satisfy (with sign convention: $u$ is positive and $v$ and $f$ are negative for convex mirror 
surface):

\begin{equation}
 \frac{1}{u}+\frac{1}{v}=\frac{1}{f}
\end{equation}
From (2.1), it is seen that the equation remains the same if $v\rightarrow u$ and $u\rightarrow 
v$. Thus, according to the principle of reversibility of light path, a real source placed at 
an object distance $u(<f)$ in front of a $concave$ mirror, has its virtual image formed at a 
distance $v$ behind the mirror. Therefore, (2.1) is called conjugate foci relationship.

Also, the height of the image $h_{i}$ in terms of the height of the object $h_{o}$ can be 
calculated from the linear magnification formula:
\begin{equation}
|\frac{h_{i}}{h_{o}}|=\ m=\frac{v}{u}
\end{equation}

\subsection{Grounded conducting sphere revisited}
\begin{figure}[h]
\begin{center}
\scalebox{0.60}{\includegraphics{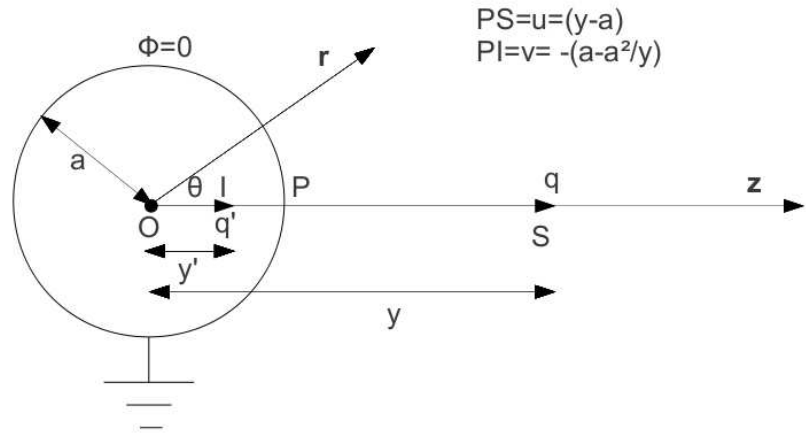}}
\caption{\label{fig:2} Grounded conducting sphere image problem}
\end{center}
\end{figure}
\subsubsection{Existence of a mirror formula} 
Referring to Figure 2, We shift the origin to the point $P$ (Pole), and define $u=y-a$ and $v
=a-\frac{a^2}{y}=\frac{a}{y}(y-a)$. Let us define the $focal \ length$ with respect to pole P 
as equal to the image charge distance when the real charge is at infinity. Notice that if $ y
\rightarrow\infty$ and hence, $u\rightarrow\infty$, we have $v\rightarrow a$). Therefore, the 
focal length in the context of the image problem is equal to radius $a$. Then, one can verify 
the following:

\begin{equation}
 \frac{1}{y-a}+\frac{1}{-\frac{a}{y}\ (\ y-\ a)}=\frac{a-y}{a(y-a)}=-\frac{1}{a}
\end{equation}
where for this convex mirror surface, object charge distance is $u=+(y-a)$ (positive) when image charge 
distance is $v=-(a-\frac{a^{2}}{y})$ (negative) and the focal distance is $f=-a$ (negative). Thus, like 
geometrical optics, in image problems (electrostatics) also there exists an analogous mirror equation:

\begin{equation}
 \frac{1}{u}+\frac{1}{v}=\frac{1}{f}=\frac{1}{a}
\end{equation}
\subsubsection{Linear Magnification formula}
The value of image charge can be simply found by using linear magnification formula as follows: 
from the boundary condition that the potentials due to real charge $q$ and the image charge $q'
$ produce sum up to zero at $P$ (Figure 2), we have:

\begin{equation}
 \frac{q'}{q}=-\frac{v}{u}=-\frac{a(y-a)}{y(y-a)}=-\frac{a}{y}
\end{equation}

Thus, $q'=-\frac{aq}{y}$ -which is correct.

\section{Visualization of field lines} 
We know that an appropriate image charge replaces the charge induced on the surface. In the context 
of our present problem, how do the resultant electric field lines look like? The answer is provided 
by Figure 3. The Mathematica plot shows the electric field lines for the charge configuration: $q=2
Q$ and $q'=-Q$, the pole being at the midpoint of $q$ and $q'$. The spherical equipotential surface 
having $\Phi=0$ appears in place of the spherical surface of grounded conducting sphere. The role of 
the rays in geometrical optics is played by lines of electric field lines in the sense that these 
field lines converge to the image charge behind the conducting surface that behaves as a mirror.
\begin{figure}[h]
\begin{center}
\scalebox{0.50}{\includegraphics{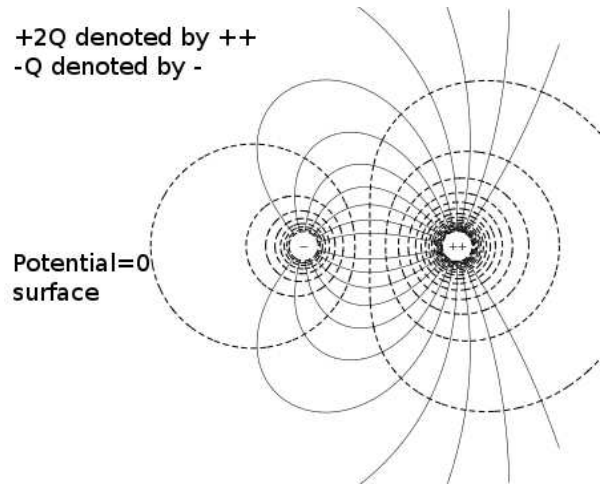}}
\caption{\label{fig:3} Field distribution for two charges in the equivalent problem}
\end{center}
\end{figure}

\section{Observations}

\subsection{Infinite grounded conducting plane image}
From (2.4), it is seen that in the limit radius tends to infinity ($a\rightarrow\infty$) the image 
distance becomes $|v|=|-\ u|=d$(say) and from (2.5) the value of the image charge becomes $q'=-q$ 
(as $a\rightarrow y$ in this limit). This analogy with optics justifies the observation that the 
grounded infinite conducting plane problem in electrostatics can be deduced from grounded conducting 
sphere problem in the limit $a\rightarrow\infty$.

\subsection{Conjugate foci relationship} 
From the form of (2.4), it is apparent that the conjugate foci relationship also holds here. How 
to see it is indeed the case? One can show that the field inside a hollow conducting sphere of 
radius $a$ (containing a point charge $Q$ at a distance $b$ from the centre) is the same as if 
there is no sphere and a charge $Q'=-\frac{aQ} {b}$ is at distance $\frac{a^2}{b}$ on the same 
axis outside the sphere. If in this inverse problem, we insist $Q\rightarrow q'=-\frac{aq}{y}$ 
and $b\rightarrow y'=\frac{a^2}{y}$, then value of the image charge becomes $Q'=-\frac{aQ}{b}=
+\frac{a\frac{aq}{y}}{\frac{a^2}{y}}=q$ -which is the value of the real charge in the original 
problem. Similarly, the distance where the image charge $Q'$ forms is $\frac{a^2}{b}=\frac{a^2
}{\frac{a^2}{y}}=y$ -which is the distance from the centre where the real charge is placed in 
the original problem. The corresponding case in optics is that a real object placed in front of 
a concave mirror (within the focal length) forms a virtual image behind the mirror.

\section{Application: Image problem for extended charge distribution}
The similarity with optics can be exploited to calculate the potential function due to an extended 
real charge distribution placed outside a grounded conducting sphere (see Figure 4). Now, invoking 
the differential form of (2.5), we have (symbols carrying the usual meaning):
\begin{equation}
\frac{dq}{u}=-\frac{dq'}{v}
\end{equation}
\begin{figure}[h]
\begin{center}
\scalebox{0.40}{\includegraphics{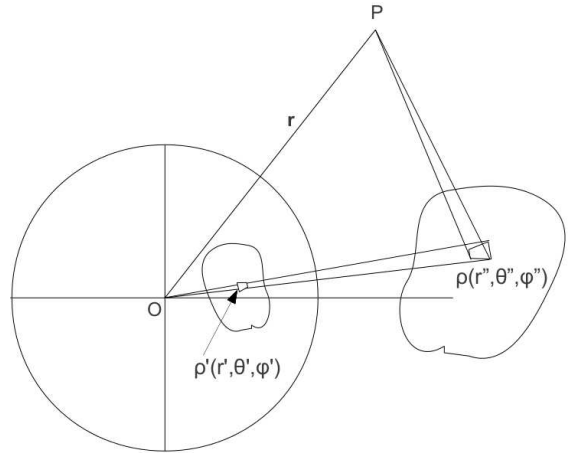}}
\caption{\label{fig:4} Image Problem for Continuous Real Charge Distribution}
\end{center}
\end{figure}
In Figure 4, we denote source and image charge distributions with double primes and single primes respectively. 
If the real charge distribution is specified with respect to the origin $O$, (5.1) becomes:
\begin{equation}
\frac{\rho(r'',\theta'',\phi'')\ d\tau''}{r''-a}=-\frac{\rho'(r',\theta',\phi')\ d\tau'}{(a-\frac{a^2}{r''})}
\end{equation}
Writing out the volume elements explicitly and noticing that the solid angle elements are equal,
\begin{equation}
\frac{\rho(r'',\theta'',\phi'')\ r''^2 dr''}{r''-a}=-\frac{\rho'(r',\theta',\phi')\ r'^2 dr'}{(a-\frac{a^2}{r''})}
\end{equation}
Notice that $r'=\frac{a^2}{r''}$ and from (2.4), the `longitudinal magnification' is $\frac{dv}{du}=-\frac{v^2}{u^2}$
\begin{equation*}
|\frac{dr''}{dr'}|=-\frac{du}{dv}=+\frac{u^2}{v^2}=\frac{(r''-a)^2}{(a-\frac{a^2}{r''})^2}
\end{equation*}
Hence, we can express the charge density of the image distribution in terms of the charge density 
of the real charge distribution as follows:
\begin{equation}
\rho'(r',\theta',\phi')=-\rho(r'',\theta'',\phi'')\frac{r''^2}{(r''-a)}(a-\frac{a^2}{r''})\frac{(r''-a)^2}{(a-\frac{a^2}{r''})^2}\frac{1}{(\frac{a^2}{r''})^2}
\end{equation}
After simplification, this becomes 
\begin{equation}
\rho'(r',\theta',\phi')=-\rho(r'',\theta'',\phi'')\frac{r''^5}{a^5}
\end{equation}
This derivation is alternative to the one given in reference [3]. However, our approach is more intuitive 
and helpful to undergraduate students. The potential at some field point $\bf{r}$ then becomes:
\begin{equation}
\Phi({\bf{r}})=\int_{real}\frac{\rho({\bf{r''}})}{|{\bf{r}}-{\bf{r''}|}}\ d\tau''+\int_{image}\frac{\rho'({\bf{r'}})}{|{\bf{r}}-{\bf{r'}}|}\ d\tau'
\end{equation}
If the boundary of the real charge distribution is known: $r''=r''(\theta,\phi)$ is known (it should be among the
known quantities of the problem), then the first integral can be evaluated. Again, given $r''=r''(\theta,\phi)$, 
we can also evaluate the boundary of the image distribution as we know $r'=\frac{a^2}{r''}$ and $\theta''=\theta'$ 
and $\phi''=\phi'$. Then, (5.6) reduces to:
\begin{equation}
\Phi({\bf{r}})=\int_{real}\frac{\rho({\bf{r''}})}{|{\bf{r}}-{\bf{r''}|}}\ d\tau''-\int_{image}\frac{r''^5}{a^5}\frac{\rho({\bf{r''}})}{|{\bf{r}}-\frac{a^2}{r''}{\hat{\bf{r''}}}|}\ d\tau'
\end{equation}
Expressing $d\tau'$ in terms of $d\tau''$, we get
\begin{equation}
\Phi({\bf{r}})=\int_{real}\frac{\rho({\bf{r''}})}{|{\bf{r}}-{\bf{r''}|}}\ d\tau''-\int_{real}\frac{r''^5}{a^5}\frac{\rho({\bf{r''}})}{|{\bf{r}}-\frac{a^2}{r''}{\hat{\bf{r''}}}|}\frac{a^6}{r''^6}\ d\tau''
\end{equation}
In the last step, the limits of the image integral has been transformed as following:
\begin{equation*}
\int_{r'_{1}}^{r'_{2}}\rightarrow\int_{\frac{a^2}{r'_{1}}}^{\frac{a^2}{r'_{2}}}\equiv\int_{r''_{1}}^{r''_{2}}
\end{equation*}
So that the expression for potential at a field point becomes:
\begin{equation}
\Phi({\bf{r}})=\int_{real}\frac{\rho({\bf{r''}})}{|{\bf{r}}-{\bf{r''}|}}\ d\tau''-\int_{real}\frac{a}{r''}\frac{\rho({\bf{r''}})}{|{\bf{r}}-\frac{a^2}{r''}{\hat{\bf{r''}}}|}\ d\tau'' 
\end{equation}
Is this result consistent with the point charge case? Let us take a point charge at a distance of $y$ 
from the centre of the conducting sphere, so that $\rho({\bf{r''}})=q\delta(\bf{r''}-\bf{y})$. Hence, 
(5.9) shows that the potential at any field point $\bf{r}$ will be 
\begin{equation*}
\Phi({\bf{r}})=\int_{real}\frac{q\delta(\bf{r''}-\bf{y})}{|{\bf{r}}-{\bf{r''}|}}\ d\tau''-\int_{real}\frac{a}{r''}\frac{q\delta(\bf{r''}-\bf{y})}{|{\bf{r}}-\frac{a^2}{r''}{\hat{\bf{r''}}}|}\ d\tau'' 
\end{equation*}
This simplifies to 
\begin{equation*}
\Phi({\bf{r}})=\frac{q}{|{\bf{r}}-{\bf{y}|}}-\frac{a}{y}\frac{q}{|{\bf{r}}-\frac{a^2}{y}{\hat{\bf{y}}}|}
\end{equation*}
-which is the familiar result. Diving by $q$, we obtain the Green's function of the problem.
\subsection{Working Principle}
In general, we have to deal with an extended continuous charge distribution. Let us say, the 
real charge density $\rho(r'',\theta'',\phi'')$ is distributed in a sphere at some fixed distance $r''
_0$ from the centre of the grounded sphere. It subtends an angle of $2cos^{-1}\frac{a}{r''_0}$ at $O$. 
Regarding the principal axis as the reference axis in the polar coordinates, the polar equation of the 
sphere is $r''^2+r''{_0}^2-2r''r''_{0}cos\theta=a^2$. For $\theta\leq cos^{-1}\frac{a}{r''_0}$, the two 
points where the real-source coordinates cut the sphere are given by the following $r''=r''(\theta'',
\phi'')$ form: $r''=r''_{0}cos\theta\pm\sqrt{a^2-{r''_{0}}^2 sin^{2}\theta}$. With this, it is possible 
to solve the two integrals in (5.9) and thereby, solve the image problem for an extended, continuous 
charge distribution. Needless to say, this form is consistent with the usual form derived in standard 
textbooks (reference [2]) for a sphere the potential on which is zero. 

\subsection{Comments}
We have made a nice guess about how to extract all information about image charge in grounded conducting 
sphere image problem in electrostatics. In return, it shed considerable light on the analogy between the 
electrostatic image and image in mirror-optics. The corollary that standard results apply equally if the 
real charge is placed inside the conducting sphere (and the image charge is produced outside the sphere) 
is just the mere reflection of the `conjugate foci' relationship in electrostatics.

The new formulation allowed us to devise the formula needed to solve the image problem for extended real 
charge near a conducting sphere. We offer a picture more vivid as well as more informative. The standard 
approach does not possibly let one know much about the distorted image charge distribution: how does its 
charge density vary or how does its boundary look like (correlation with the boundary of the real charge). 
In comparison, our treatment directly finds these details and eventually, just add the potentials due to 
real charge and image charge, as is done for the point charge case.

By the way, if the potential on the conducting sphere is non-zero, Dirichlet boundary condition is to be 
used. This results in the following additional term in our formula: 
\begin{equation*}
-\frac{1}{4\pi}\oint\Phi(a,\theta'',\phi'')\frac{(r^2-a^2)}{a(r^2+a^2-2ar cos\gamma)^{3/2}}a^2 d\Omega''
\end{equation*} 
where $cos\gamma=\hat{\bf{r}}\cdot\hat{\bf{r''}}$. In the standard approach also, one needs to add 
this boundary term to the generic Green's function for the `grounded' conducting sphere.

We wish to remind the reader that the analogy we have seen in this article is not complete. In optics, a 
spherical mirror can be a portion of a sphere and still can form an image. Because only those light rays 
that are close to the axis are needed to form it. However, in electrostatics, one needs a full spherical 
conducting surface to form the image. This also explains why do we need infinite plane conductor to form 
an electrostatic image but a finite plane mirror suffices to form optical image. We chose the term mirror 
(and not the term lens) as the electric field $\bf{E}$ of the real charge does not penetrate the conductor 
like light ray actually does not penetrate a mirror. 

\section{Pedagogic Interest of the article}
Image problems are part of the physics curriculum in undergraduate and graduate level courses in 
universities. To the undergraduate students in introductory electrodynamics course, the analogy 
given in section $2$ (mirror equation, magnification formula and conjugate foci relation) will be 
a pleasant surprise and a more familiar way to calculate image charges (as they are expected to 
have covered lens/mirror optics in high school level). On the other hand, the graduate students 
will find the alternative treatment of image problem for extended charge distribution interesting. 
The article deals with an interdisciplinary topic; so, it might attract the general physicists as 
well while they see that the same basic principle connects apparently different areas.

\ack
I am very much grateful to sincere supports and encouragements from Dr. Debapriyo Syam, Sayan, 
Tanmay, Manoneeta and Tamali. The anonymous reviewer of the paper also helped me in improving 
the article.

\end{document}